\documentclass[
]{scrarticle}

\usepackage{amsmath, amssymb}
\usepackage{graphicx,color}
\usepackage[
  backend=biber,
  style=numeric,
  sorting=none
]{biblatex}

\usepackage{authblk}
\usepackage{orcidlink}

\newcommand{\br}[1]{\left( #1 \right)}

\newcommand{\mathdefault}[1][]{} 

\newcommand{\ket}[1]{\vert {#1}\rangle}

\newcommand{\matrixel}[3]{\langle #1 \vphantom{#3} | #2 | #3 \vphantom{#1} \rangle}

\newcommand{\red}[1]{{\color[rgb]{0.7,0,0} #1}}

\makeatletter
\newcommand{\hrules}{%
\@for\i:=1,2,3,4,5\do{%
\noindent\red{\rule{\linewidth}{1pt}}\par\vspace{-1ex}%
}%
\vspace{1ex}
}
\makeatother

\newcommand{\tf}{t_\mathrm{f}}

\newcommand{\xii}{\xi^{(r)}}

\newcommand{\sz}[1][]{\ifthenelse{\equal{#1}{}}{\sigma_z}{\sigma_z^{(#1)}}}
\newcommand{\sx}[1][]{\ifthenelse{\equal{#1}{}}{\sigma_z}{\sigma_x^{(#1)}}}
\newcommand{\sm}[1][]{\ifthenelse{\equal{#1}{}}{\sigma_-}{\sigma_-^{(#1)}}}
\renewcommand{\sp}[1][]{\ifthenelse{\equal{#1}{}}{\sigma_+}{\sigma_+^{(#1)}}}

\renewcommand{\sz}[1][]{\ifthenelse{\equal{#1}{}}{\sigma^z}{\sigma^z_{#1}}}
\renewcommand{\sx}[1][]{\ifthenelse{\equal{#1}{}}{\sigma^x}{\sigma^x_{#1}}}
\renewcommand{\sm}[1][]{\ifthenelse{\equal{#1}{}}{\sigma^-}{\sigma^-_{#1}}}
\renewcommand{\sp}[1][]{\ifthenelse{\equal{#1}{}}{\sigma^+}{\sigma^+_{#1}}}

\newcommand{\rhof}{\rho_\mathrm{f}}
\usepackage{listings}
\begin{document}

\title{Testing Noise Correlations by an AI-Assisted Two-Qubit Quantum Sensor}

\author[1,2]{Dario Fasone\thanks{dario.fasone@dfa.unict.it}\orcidlink{0009-0002-4219-9671}}
\author[1]{Shreyasi Mukherjee\thanks{shreyasi.mukherjee@dfa.unict.it}}
\author[3]{Mauro Paternostro\orcidlink{0000-0001-8870-9134}}
\author[1,4]{Elisabetta Paladino\orcidlink{0000-0002-9929-3768}}
\author[1,4]{Luigi Giannelli\orcidlink{0000-0001-9704-7304}}
\author[1,4]{Giuseppe A.~Falci\thanks{giuseppe.falci@unict.it}\orcidlink{0000-0001-5842-2677}}

\affil[1]{Dipartimento di Fisica e Astronomia ``Ettore Majorana'', Università di Catania, Italy}
\affil[2]{PhD in Quantum Technologies, Università di Napoli Federico II, Italy}
\affil[3]{Dipartimento di Fisica e Chimica ``Emilio Segré'', Università di Palermo, Italy}
\affil[4]{INFN, Sezione di Catania, Italy}

\renewcommand\Authands{, }
\setlength{\affilsep}{0.5em}
\renewcommand\Affilfont{\small}

\date{}
\maketitle
\begin{abstract}
 We introduce and validate a machine learning-assisted protocol to classify time and space correlations of classical noise acting on a quantum system, using two interacting qubits as probe. We consider different classes of noise, according to their Markovianity and spatial correlations. Leveraging the sensitivity of a coherent population transfer protocol under three distinct driving conditions, the various noises are discriminated by only measuring the final transfer efficiencies. This approach reaches around 90\% accuracy with a minimal experimental overhead.
\end{abstract}
    The interaction with environmental degrees of freedom makes quantum hardware prone to   decoherence~\cite{ZurekRMP2003decoherence} which would erase all the advantages of quantum coherence. While \emph{single} qubits are nowadays well optimized protected from decoherence~\cite{kjaergaard_superconducting_2020}, substantial work has to be done in upscaled quantum architectures where, in particular, effects of time-correlated~\cite{PaladinoRMP2014$1,falci_1f_2024} and space-correlated~ noise were analyzed~\cite{darrigo_effects_2008,PaladinoRMP2014$1,zou_spatially_2023} and detected~\cite{bylander_noise_2011,von_lupke_two-qubit_2020,boter_spatial_2020,rojas-arias_spatial_2023,YonedaNP2023noisecorrelation}. 
    Space correlations of non-Markovian noise directly affect two-qubit gates built on the Ising-$xx$ interaction~\cite{darrigo_open-loop_2024} and quantum error correction~\cite{li_quantum_2023}. 
    Therefore, new methods for noise-diagnostics and strategies to mitigate its effects are paramount for advances in quantum technologies.noise may emerge
    
 In this work, we propose a design of a quantum sensor for testing the presence of time- and space-correlated noise in solid-state quantum hardware, avoiding direct measurement of the noise cross-spectra. The principal system consists of two ultrastrongly coupled qubits, their coupling strength $g$  being comparable to the individual Bohr energies $\sim \epsilon$. It may be a quantum sensor detecting material properties of a substrate or a subsystem of a larger quantum processing unit. We start with the Hamiltonian ($\hbar = 1$) \begin{equation}
    H_{\mathrm sys} = -\frac{\epsilon}{2}\sigma_1^z -\frac{\epsilon}{2}\sigma_2^z +\frac{g}{2}\sigma_1^x\sigma_2^x,
\end{equation}
   with eigenvectors $\{\ket{0},\ket{1},\ket{2},\ket{3}\}$ and eigenvalues $\{-\varepsilon,\varepsilon,\tfrac{g}{2},-\tfrac{g}{2}\}$, where $\varepsilon = \tfrac{1}{2}\sqrt{\epsilon^2 + g^2}$.
The system is subject to local longitudinal noise, which induces fluctuations of the individual qubit splittings, modeled by two classical stochastic processes~\cite{Mandel1995optical} $\delta_i(t)$ 
 \begin{equation}
    H_{\mathrm noise}(t) = -\frac{\delta_1}{2}\sigma_1^z -\frac{\delta_2}{2}\sigma_2^z,
\end{equation} 
We consider three non-Markovian and two Markovian noise classes:
\begin{itemize}
    \item \textbf{Non-Markovian:} we consider the limit of quasistatic noise where $\delta_i$ are random variables picked from a Gaussian distribution. We identify three distinct classes: correlated, anticorrelated, and uncorrelated local variables $\delta_i$.
    \item \textbf{Markovian noise:} zero-mean, delta-correlated stochastic processes. We consider the correlated and anticorrelated local processes.
\end{itemize}
The system is operated to produce coherent population transfer through a STIRAP-like protocol~\cite{VitanovRMP2017stimulated,falci_design_2013,giannelli_tutorial_2022}. Looking for a suitable design, we first consider the driven system in the absence of noise. Favorable conditions are found by operating in the ultra-strong coupling regime $g \sim \epsilon$, and by driving symmetrically the two qubits. 
\begin{equation}
    H_c(t) = W(t)\,\br{\sigma_1^x + \sigma_2^x},
\end{equation}
This symmetry enforces a selection rule that excludes $\ket{3}$ from the dynamics, which is then limited to a three-level system. Control is operated by a two-tone field,  $W(t) = \Omega_{20}(t)\cos(\omega_{20} t) + \Omega_{12}(t)\cos(\omega_{12} t)$, where $\Omega_{ij}(t)$ are slowly-varying pulse envelopes and $\omega_{ij}$ match the energy splittings between eigenvalues $i-j$. In a doubly rotating frame and after using the rotating wave approximation (RWA) for $H_c$  we obtain the Hamiltonian
\begin{equation}
\tilde{H}=\frac{1}{\sqrt{2}}\{\Omega_p(t)\lvert0\rangle\langle2\rvert+\Omega_s(t)\lvert1\rangle\langle2\rvert+\mathrm{h.c.}\}.
\end{equation}
which implements a ladder configuration. Then coherent population transfer by STIRAP can be obtained using a suitable time dependence of pulse envelopes $\Omega_{p/s}(t)$.  

Asymmetries and imperfections, such as those caused by noise, modify this picture, since coherence is suppressed and selection rules are relaxed. 
The resulting 4-level dynamics, while deteriorating the efficiency of population transfer, has the key advantage of yielding an increased discrimination between different classes of noise correlations.
\begin{figure}[t!]
    \centering
    \includegraphics[width=0.8\textwidth]{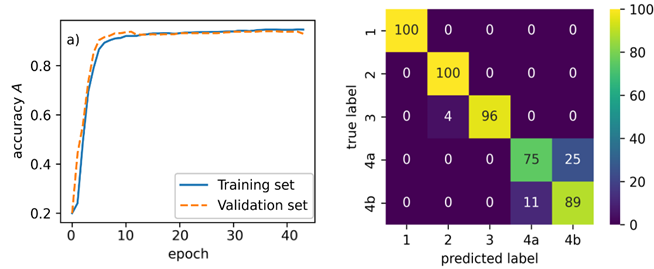}
    \caption{(a) Accuracy for the training (solid blue line) and validation (solid orange line) sets versus the number of epochs. (b) Confusion matrix of the model for classifying the noise types. Each row represents the true noise class, while each column corresponds to the predicted class. All five noise classes can be classified with accuracy up to $\sim 92\%$.}
    \label{fig:accuracy}
\end{figure}
To utilise the most accessible measurement protocol, we employ, as the figure of merit for the Neural Network, the average population of the state $\ket{ee}$, that is \begin{equation}
        \xi=\lim_{N\to\infty}\frac{1}{N}\sum_{r=1}^{N}\xi^{(r)}
    \end{equation}
    where $\xii=\matrixel{ee}{\rhof^{(r)}}{ee}$ and $\rhof^{(r)}$ is the density
    matrix of the system at the final time $\tf$ for the $r$-th noise realization. Such quantity is computed accordingly for each of the noise classes, under 3 driving conditions, $
(i)\Omega_p^{\max}=\Omega_s^{\max},(ii)\Omega_p^{\max}=2\Omega_s^{\max},(iii)\Omega_p^{\max}=\Omega_s^{\max}/2$.
We use synthetic data generated by numerical simulations. For the correlated classes the correlation parameter is randomly drawn, while for the uncorrelated classes, the Gaussian width of the noise distributions is varied. For each class, we obtained 500 data points, each consisting of a 3-dimensional vector containing the average efficiency for the fixed noise parameters under the 3 driving conditions.

The dataset is used to classify noise affecting the qubits by Supervised Learning. The training~\cite{MarquardtSPLN2021machine} is performed by minimizing the sparse categorical cross-entropy, which measures the distance between the predicted label and the true label of the noise. The model reaches an accuracy of around $92\%$ (Fig.~\ref{fig:accuracy}a). The accuracy for the test datasets is summarised in Fig.~\ref{fig:accuracy}b.

The model achieves an accuracy of 99.4\% in distinguishing between non-Markovian and Markovian noise. Within the non-Markovian noise class, it correctly classifies correlations with an accuracy of 98.67\%, whereas within the Markovian class, the classification accuracy is 82\%. This contrasts with the three-level system case analyzed in~\cite{MukherjeeMLST2024noise}, where the model was unable to discriminate between the two distinct Markovian noise types. 

We finally observe that current experiments on time-\cite{bylander_noise_2011} and space- ~\cite{von_lupke_two-qubit_2020,boter_spatial_2020,zou_spatially_2023,rojas-arias_spatial_2023,YonedaNP2023noisecorrelation} correlations characterize noise via the 
measurement of power spectra and cross spectra which is a highly demanding procedure, very hard to scale to larger quantum structures. Instead, the procedure we propose aims at detecting global properties of noise, as the existence of correlations, irrespective on their detailed form. 

\section*{Acknowledgments}
DF and LG acknowledge support from the PNRR MUR project PE0000023-NQSTI "National Quantum 
Science and Technology Institute" - Spoke 1; SM acknowledge support from the "Centro Nazionale di Ricerca in High-Performance Computing, Big Data and Quantum Computing"-ICSC; GF acknowledges support from PRIN 2022 "SuperNISQ"; GF and EP acknowledge support from the University of Catania, Piano Incentivi Ricerca di Ateneo 2024-26, project
QTCM; EP acknowledges the COST Action SUPERQUMAP (CA 21144); MP acknowledges support from the European Union’s Horizon 
Europe EIC-Pathfinder project QuCoM (101046973), the Department for the Economy of Northern Ireland under the US-Ireland R\&D Partnership Programme, and the PNRR MUR project PE0000023-NQSTI - SPOKE 2 through project ASpEQCt.

\section*{Declaration on Generative AI}
  
  The authors have not employed any Generative AI tools.
\printbibliography

\end{document}